# Mechanically Tunable Slippery Behavior on Soft Poly(dimethylsiloxane) (PDMS) Based Anisotropic Wrinkles Infused with Lubricating Fluid

Pritam Kumar Roy,[a] Reeta Pant,[a] Arun Kumar Nagarajan,[b] Krishnacharya Khare[a*]

[a]Department of Physics, Indian Institute of Technology Kanpur, Kanpur - 208016, India

[b]Hindustan Unilever Research Centre, Bangalore - 560066, India

Email: kcharya@iitk.ac.in

**KEYWORDS:** Slippery, Wrinkle, Tunable topography, Wetting, Liquid-liquid interface

## Abstract

We demonstrate a novel technique to fabricate mechanically tunable slippery surfaces using one dimensional (anisotropic) elastic wrinkles. Such wrinkles show tunable topography (amplitude) on the application of mechanical strain. Following Nepenthes pitcher plants, lubricating fluid infused solid surfaces show excellent slippery behavior for test liquid drops. Therefore combining the above two i.e. infusing suitable lubricating fluid on elastic wrinkles would enable us to fabricate mechanically tunable slippery surfaces. Completely stretched (flat) wrinkles have uniform coating of lubricating fluid whereas completely relaxed (full amplitude) wrinkles have most of the lubricating oil in the wrinkle grooves. Therefore water drops on completely stretched surface show excellent slippery behavior whereas on completely relaxed surface they show very poor slippery behavior. Therefore continuous variation of wrinkle stretching provide reversibly tunable slippery behavior on such system. Since the winkles are one dimensional, they show anisotropic tunability of slippery behavior depending upon whether test liquid drops slip parallel or perpendicular to the wrinkles.

## Introduction

Elastic winkles with mechanically tunable topography have been shown as one of the best candidates in many applications including tunable adhesion, wetting, microfluidics, microlens array, optical grating etc.[1-10] Such wrinkles depict the above mentioned tunable properties as their amplitude or height can be reversibly tuned as a function of applied external strain. Fabrication of such wrinkles on elastic substrates have been demonstrated in various ways for different experimental systems.[4, 11-29] Buckling instability in a bilayer system with elastic modulus mismatch leads to the spontaneous wrinkle formation.[12-13, 30] Resulting wrinkle can be isotropic or anisotropic depending upon the nature of external strain (1 or 2 dimensional).[19, 31] Crosslinked elastomeric polymer polydimethylsiloxane (PDMS) has commonly been used as thick elastic foundation (substrate) in most of the wrinkle fabrications. Expanding of the substrate layer is often achieved by mechanical stretching, heating, swelling etc. Subsequently a thin rigid layer is deposited on top of the stretched elastic substrate followed by relaxation which spontaneously generates the wrinkles. For small strains, the wavelength ($\lambda_0$) and amplitude ($A_0$) can be derived using linear buckling theory as: $\lambda_0 = 2\pi t(\bar{E}_f/3\bar{E}_s)^{1/3}$ and $A_0 = t((\varepsilon_0/\varepsilon_c) - 1)^{1/2}$ where $\varepsilon_0$ is applied strain, $\varepsilon_c$ is the critical strain given as $\varepsilon_c = -1/4\,(3\bar{E}_s/\bar{E}_f)^{2/3}$, $\bar{E} = (E/1 - v^2)$, $E$ is elastic modulus, $v$ is Poisson's ratio and $f$ and $s$ represent film and substrate respectively.[32-36] Resulting wrinkle's topography (wavelength, amplitude) can be controlled by manipulating the elastic properties of the substrate and film, thickness of the film and applied strain.

Aizenberg et al. demonstrated lubricant infused slippery surfaces on porous Teflon membranes inspired by Nepenthes pitcher plants.[37-38] Such slippery surfaces are found very useful in many applications including self cleaning, drag reduction, anti-icing, enhanced condensation etc.[37-43] Basic requirement of slippery surfaces is: (i) lubricating fluid should be completely wetting on the substrate, (ii) test (slipping) liquid should be immiscible with the lubricating fluid, and (iii) the test liquid should be non-wetting on the substrate.[37-38, 44-45] For hydrophilic substrates, water drops sink in the lubricating layer thus does not show slippery behavior.[37, 44] Therefore slippery surfaces for water (test liquid) can be fabricated using any oil and hydrophobic substrates. Various research groups used different fabrication techniques to fabricate lubricating fluid infused slippery surfaces. AIzenberg et al. used porous Teflon membrane and multi layered silica nanoparticles with silane functionalization with fluorinated lubricating oil. Such surfaces showed excellent slippery behavior for various test liquid e.g. water, glycerol, ethylene glycol, alkanes, biofluids, crude oil etc.[37] Varanasi et al. used lithographically patterned micro/nanotextured hydrophobic substrates infused with silicone oil to slip aqueous drops.[45-47] Recently Zhang et al. demonstrated anisotropic sliding of water droplets on anisotropic micro-grooved organogel surfaces infused with silicone oil as lubricating fluid.[48] They showed that water drops find it easy to slide along groove directions compared to the direction perpendicular to the grooves. Since the micro-grooves used in the experiment were fabricated using a template, they could only demonstrate qualitatively the stretching dependent sliding of water droplets. Therefore combining the lubricating fluid infused slippery surfaces with mechanically tunable elastic wrinkles would enable us to fabricate mechanically tunable slippery surfaces with better control and reversibility.

In this paper, we here present a novel technique to fabricate mechanically tunable slippery surfaces using lubricating fluid infused PDMS based anisotropic wrinkles. One dimensional linear wrinkles coated with silicone oil lubricating fluid provide anisotropic slippery behavior in parallel and perpendicular direction which depends upon applied external strain.

Contact angle hysteresis, critical tilt angle for slippage and slip velocity of water drops at constant tilt angle were used as experimental parameters to demonstrate the slippery behavior as a function of applied strain.

**Experimental Section**

Flat PDMS sheets were prepared using Sylgard® 184 (silicone elastomer) by mixing the base and cross linker in 10:1 ratio and sandwiching it between two glass plates (separated by 1.5 mm thick spacers) followed by curing at 85°C for 3 hours. The spacers thickness decide the thickness of the resulting PDMS sheets which in our case was kept constant at 1.5 mm. Crosslinked PDMS was taken out of the oven and the glass plates were peeled out.

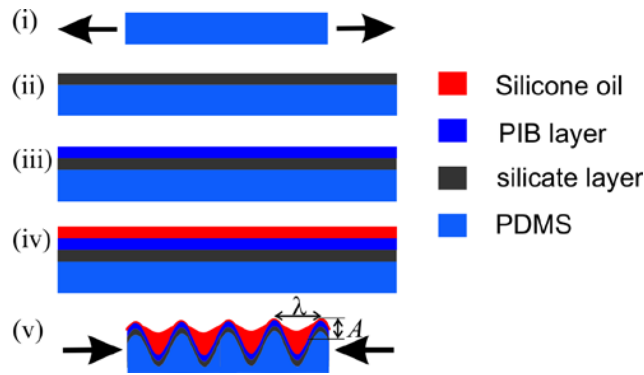

Figure 1: Schematics of wrinkle based tunable slippery surface fabrication; (i) model PDMS substrate, (ii) UVO exposure to stretched PDMS substrate for oxidation, (iii) coating the surface with PIB hydrophobic layer, (iv) coating silicone oil lubricating fluid, and (v) releasing the applied strain to generate wrinkles coated with lubricating fluid.

PDMS sheets were then cut into 3×2 cm$^2$ pieces and used as substrates for all the experiments. To fabricate wrinkles, rectangular PDMS sheets were clamped in a homemade stretching device, stretched to 50% strain along the long axis and exposed to UV ozone (UVO) (144AX, Jelight

Company) for 60 mins. Top surface of PDMS sheet gets oxidized during UVO exposure forming dense silicate layer which spontaneously results into wrinkles upon releasing the applied strain. Due to top silicate layer, fabricated wrinkles were hydrophilic which were made hydrophobic by coating a thin layer of polyisobutylene (PIB). Wrinkle morphologies were analyzed using a scanning electron microscope (SEM) (Quanta 200, FEI) and 3D optical profilometer (NanoMap D, AEP Technology). Silicone oil (350 cSt) was used as lubricating fluid and was dip coated on completely stretched wrinkles followed by releasing the applied strain. Schematic of complete wrinkle fabrication process and lubricant coating is summarized in Figure 1. Slippery experiments were performed by measuring contact angle hysteresis (contact angle difference between advancing and receding drop volume cycle), critical tilt angle (minimum tilt angle at which drops start slipping) and slip velocity of water drops (20 μl volume) at constant tilt angle of 20°. The slippery measurements were done using a contact angle goniometer (OCA35, DataPhysics Germany) and upright optical microscope (BX51, Olympus). All these measurements were done along parallel and perpendicular direction of wrinkles as a function of applied external strain.

**Results and Discussion**

Upon releasing the strain of a stretched PDMS sheet after UVO exposure, wrinkles appear spontaneously. Figure 2 (a) & (b) shows SEM and 3D optical profilometer images of a wrinkle surface. Due to elastic PDMS substrate, these wrinkles are mechanically tunable that means their amplitude can be reversibly altered as a function of applied strain. Exploiting buckling instability with large strain, wavelength ($\lambda$) and amplitude ($A$) of the wrinkles can be derived as:

$$\lambda = \frac{\lambda_0(1+\varepsilon)}{(1+\varepsilon_0)(1+\varepsilon+\zeta)^{1/3}} \quad \& \quad A = \frac{t\sqrt{\frac{(\varepsilon_0-\varepsilon)}{\varepsilon_c}-1}}{\sqrt{(1+\varepsilon_0)}(1+\varepsilon+\zeta)^{1/3}} \tag{1}$$

where $\zeta = \frac{5}{32}(\varepsilon_0 - \varepsilon)(1+\varepsilon_0)$.[33] Controlling various experimental parameters, e.g. PDMS elastic modulus, pre-strain, UVO exposure time would result in winkles of different wavelength and amplitude. Wrinkles used in the experiments had pre-strain $\varepsilon_0 = 50\%$, which resulted in wavelength $\lambda = 50$ µm with corresponding amplitude $A = 11.5$ µm and was kept constant throughout the experiments. Figure 2(c) shows the tunablility of wrinkle topography (wavelength and amplitude) upon applying strain. Black and red data points are experimental values and the corresponding solid lines are theoretical values given via Eq. (1).

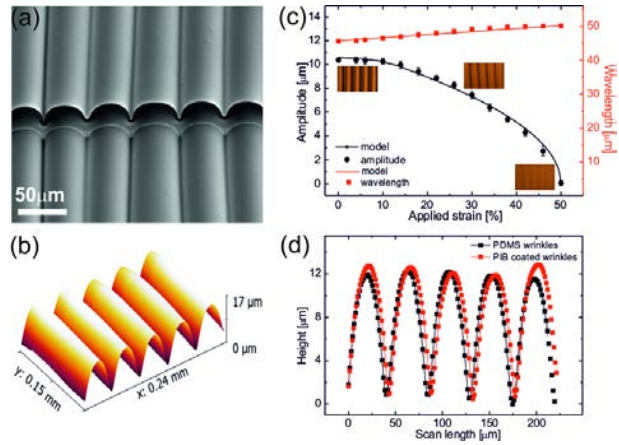

Figure 2: (a) SEM and (b) 3D optical profilometer images of wrinkle surface, (c) tunable topography (wavelength and amplitude) of wrinkles as a function of applied strain, and (d) scan line profiles of wrinkle surface before and after PIB coating showing negligible change in surface topography.

It is clear that upon applying 50% strain, the wrinkles become almost flat ($A \approx 0$) with very small increase in its wavelength. The PDMS wrinkles were coated with a thin PIB layer to provide hydrophobic surface which is essential for stable slippery surfaces. Figure 2(d) shows 3D optical profilometer scan line profiles of wrinkle surface before and after PIB coating confirming negligible change in surface topography. PIB coated wrinkles were found oleophilic and hydrophobic in nature with silicone oil and water contact angles as 8° and 111° respectively as shown in Fig. 3(a) & (b). Therefore the top PIB coating fulfils the required boundary conditions for stable slippery surfaces for water drops.[44] Figure 3(c) shows the top view of a water droplet deposited on dry wrinkles (without lubricating oil coating).

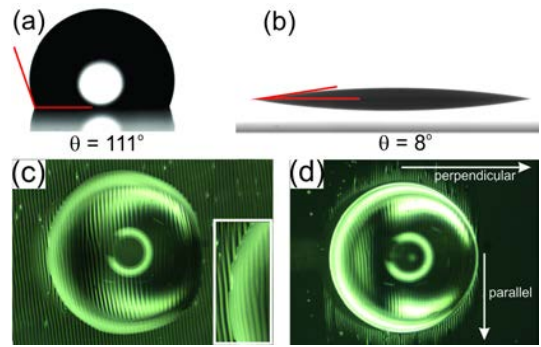

Figure 3: Optical contact angle images of (a) water and (b) silicone oil drops on a flat PIB surface showing the hydrophobic and oleophilic nature of PIB. (c) & (d) Top view of a water drop on PIB coated wrinkles before and after silicone oil coating. Inset of (c) shows zoomed in image showing pinning of three phase contact line of a water drop.

Due to the underlying topography, the three phase contact line of the drop gets pinned along the wrinkles as shown in the inset.[49] Lubricating fluid, silicone oil, was dip coated on stretched (flat) wrinkles which forms uniform thin film and its thickness was estimated to 3 μm via weight difference method. After releasing the pre-strain, wrinkles are generated and silicone oil flows to the grooves (bottom portion of wrinkles) due to capillary action which leaves extremely thin lubricant film on wrinkle crests. The lubricating fluid thickness variation at groove and crest portions of wrinkles is responsible for tunable slippery behavior which can be reversibly controlled by applied strain. Figure 3(d) shows the top view of a water droplet deposited on silicone oil coated wrinkles which shows completely circular drop shape indicating no pinning of the three phase contact line of the water drop. Perpendicular and parallel directions for water drop motion is defined with respect to the wrinkle orientation i.e. perpendicular is the direction when water drops move normal to the wrinkles and parallel is when they move along the wrinkles (cf. Figure 3(d)). Slippery behavior of silicone oil coated wrinkles were first analyzed by measuring the critical tilt angle for water drops as well as contact angle hysteresis on lubricated wrinkles as a function of applied strain. As shown in Fig. 1(v), in case of completely relaxed wrinkles, most of the lubricating fluid fills the groove portion of the wrinkles and only very thin layer of oil covers the wrinkle crests, which is also represented schematically in Fig. 4(a). As the applied strain is increased (sample is stretched), wrinkle amplitude decreases resulting in increased oil film thickness on wrinkle crests due to flow of oil from grooves. For maximum applied strain, 50% in the present case, wrinkles become completely flat resulting in almost homogeneous film of lubricating fluid on the surface. In this case the lubricating film thickness becomes equal to the as deposited initial film which in the present case is 3 μm. Therefore on wrinkle's crest, lubricating film thickness vary between almost zero to 3 μm depending upon the applied strain. Variation of lubricating oil height in wrinkle groove can be calculated via simple calculation using volume conservation of lubricating fluid. For small pre-strains (< 10%), wrinkles are sinusoidal in nature and contour length of a wrinkle can be written as: $l = \int_0^\lambda \sqrt{1 + \left[\frac{d}{dx}\left\{A \sin\left(\frac{2\pi x}{\lambda}\right)\right\}\right]^2} \, dx$ which depends on wrinkle amplitude (*A*) and wavelength (*λ*). Wrinkles generated with large pre-strains (~50%) resemble triangular grooves with sharp bottom corner as shown in Fig. 2(b). Therefore height of the lubricating oil in a wrinkle groove can be approximated as:

$$h = T + \frac{\Delta LT}{(L-\Delta L)} \qquad \text{when oil thickness is more than groove height } (T > H),$$

and

$$h = \sqrt{\frac{4A}{n\lambda}(nA\lambda - LT)} \qquad \text{when oil thickness is smaller than groove height } (T < H).$$

Here $T$ is the thickness of lubricating oil on completely stretched sample, $L$ is the length of a completely stretched sample, $\Delta L$ is the change in length due to strain and $n$ is the total number of grooves in the sample. Also the wrinkle groove height is defined as twice of its amplitude ($H = 2A$).

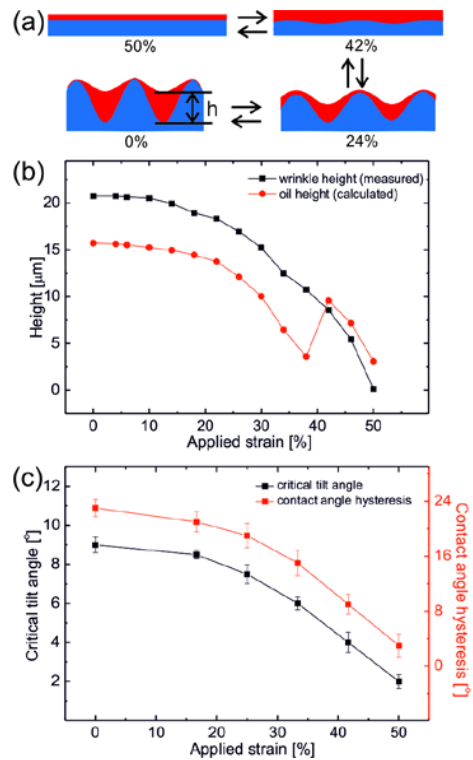

Figure 4: (a) Schematic of lubricating film thickness variation on wrinkles upon releasing the applied strain, (b) plot of the wrinkle and oil height as a function of releasing the applied strain and (c) critical tilt angle and contact angle hysteresis (in perpendicular direction) with applied strain on silicone oil coated wrinkles.

Figure 4 (b) shows plot of lubricating oil height (calculated) with wrinkle height (measured) in a wrinkle groove as a function of applied strain. As mentioned earlier, lubricating oil thickness on completely stretched (50% strain) sample was 3 μm. Releasing the applied strain up to 42%, whole wrinkled sample gets compressed which leads to the increase in oil thickness with little increase in wrinkle amplitude. Till this applied strain, since the wrinkle height is smaller than the oil height, oil layer still covers the entire sample surface. Below 40% applied strain, wrinkle amplitude become larger than the oil film thickness. This is the crossover position in the plot (Fig. 4(b)) at which the lubricating oil start moving to the wrinkle groove, leaving behind very thin oil layer at wrinkle crests. With further decreases in applied strain, wrinkle amplitude increases more with corresponding increase and decrease in oil height in wrinkle grooves and crests respectively. This whole cycle has been schematically shown in Figure 4(a). Critical tilt angle and contact angle hysteresis (in perpendicular direction) as a function of applied strain is plotted in Figure 4(c). Completely relaxed wrinkles, with 0% applied strain, depict very large critical tilt angle (~ 9°) and corresponding contact angle hysteresis (~ 23°). This is due to large amplitude of wrinkles which acts as barrier during water drop motion. Also very thin layer of silicone oil film on wrinkle crests is not sufficient to provide good lubrication for smooth motion of three phase contact line of water drops. In the reverse cycle, as applied strain is increased, wrinkle amplitude decreases which also decreases the lubricant height in grooves and increases the lubricant height on wrinkle crests. This results in decreased critical tilt angle for slippage as well as smaller contact angle hysteresis. Therefore samples are expected to show better slippery behavior with increase in applied strain. Crossover of lubricant height with wrinkle height is obtained around 42% applied strain above which lubricant height become larger than the wrinkle height. Around this point, the lubricant covers the entire wrinkle surface which results in much smaller values of critical tilt angle and contact angle hysteresis. In parallel direction, not much dependence on applied strain is seen as water drops slip along the wrinkles and do not cross them therefore do not feel any barrier.

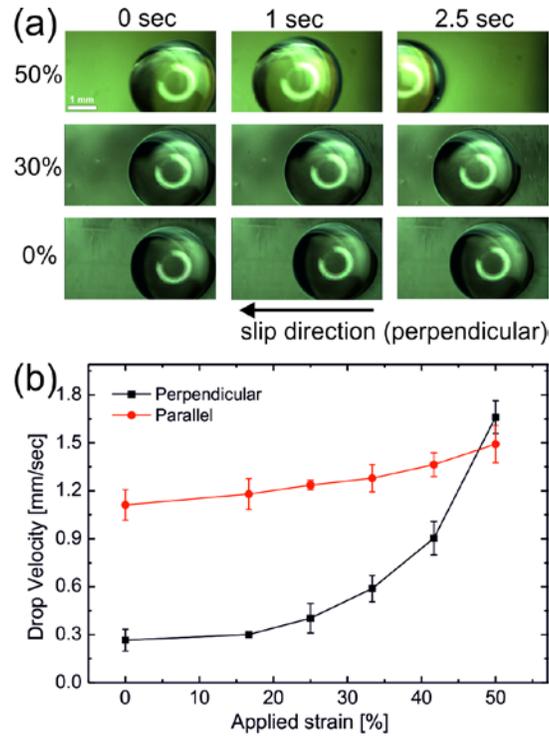

Figure 5: (a) Snap shots of a water drop slipping on silicone oil coated wrinkles with different applied strains, (b) slip velocity of water drops as a function of applied strain in parallel and perpendicular directions.

Figure 5 (a) shows optical snapshots of 20 µl water drops slipping on lubricated wrinkles at different times in perpendicular directions at 20° tilt angle for different applied strains. First row shows almost flat wrinkles, with 50% strain, showing highest velocity or lowest friction due to negligible wrinkle amplitude or barrier height. With 30% strain, due to increase in wrinkle amplitude which increases the barrier height, slip velocity is decreased as clearly shown by the lagging water drop after 2.5 sec. At 0% applied strain, full amplitude wrinkles offer maximum barrier to slipping water drops showing the lowest velocity. (see supporting Movies S1, S2 and S3 corresponding to 50% strain perpendicular, 0% strain parallel and 0% strain perpendicular respectively) Due to lubricating fluid coating, the global shape of water drops remain circular indicating no pinning of three phase contact line as also seen in Fig. 3 (d). That is why, even for the completely relaxed wrinkles with full amplitude, slip velocity of water drops is non-zero confirming no pinning. Slip velocity of water drops at different strains in both parallel and perpendicular directions is shown in Fig. 5(b). The plot clearly shows anisotropic mechanically tunable slippery behavior for water drops on lubricating fluid coated wrinkles. In perpendicular direction, the slip velocity changes with almost one order of magnitude whereas in parallel direction, not much change in slip velocity is observed with change in applied strain. Therefore lubricating oil coated 1D wrinkles are one of the best candidate to depict anisotropic as well as mechanically tunable slippery behavior for water drops.

**Conclusion:**

In summary, we demonstrated mechanically tunable slippery surfaces based on lubricating silicone oil coated anisotropic wrinkles. One dimensional wrinkles were fabricated in PDMS elastomer following standard procedure exploiting buckling instability in bilayer system. The elastic wrinkles show tunable topography depending upon applied strain. Dry wrinkles, without any lubricant, does not show any slippery behavior as three phase contact line of water drops get pinned on wrinkle surface making them sticky. Wrinkles, in completely stretched condition, were dip coated with lubricant silicone oil followed by gravity drainage to remove excess oil resulting in uniform lubricating film. Subsequently, the applied strain was released resulting in lubricating oil coated wrinkles where most of the oil goes in wrinkle grooves and very thin oil layer remains on wrinkle crests. After lubricant coating, even the completely relaxed wrinkles with full amplitude don't show any pinning as global shape of water drops remain circular and they slip with finite velocity. Wrinkle height, which can be modulated as a function of applied strain, acts as barrier for slipping water drops. With increasing applied strain, wrinkle amplitude decreases which reduces the barrier height for slipping water drops in perpendicular direction. For parallel direction, not much effect of applied strain is visible on drop velocity as slipping drops do not cross the wrinkles (barrier). Corresponding contact angle hysteresis and the critical tilt angle also decreases for water drops in perpendicular direction.

**Acknowledgements:**

This research work was supported by Hindustan Unilever Limited, India and DST, New Delhi through its Unit of Excellence on Soft Nanofabrication at IIT Kanpur.